\documentclass[%
 aip,
 jmp,%
 amsmath,amssymb,
 reprint,%
 twocolumn,
 superscriptaddress,
]{revtex4}

\usepackage{stmaryrd}
\usepackage{graphicx}
\usepackage{bm}
\usepackage{textcomp}
\usepackage{subfigure}
\usepackage{multirow,amssymb,amsbsy,amsmath}
\usepackage{ctable}
\usepackage{colortbl}
\usepackage{amsmath}

\begin{document}

\title{Detection of incoherent broadband terahertz light using antenna-coupled high-electron-mobility field-effect transistors}

\author{Hua Qin}\email[Corresponding author, ]{hqin2007@sinano.ac.cn}
\affiliation{
Key Laboratory of Nanodevices and Applications, 
Suzhou Institute of Nano-tech and Nano-bionics (SINANO), 
Chinese Academy of Sciences, 
398 Ruoshui Road, Suzhou 215123, P.~R.~China}

\author{Xiang Li}
\affiliation{
Key Laboratory of Nanodevices and Applications, 
Suzhou Institute of Nano-tech and Nano-bionics (SINANO), 
Chinese Academy of Sciences, 
398 Ruoshui Road, Suzhou 215123, P.~R.~China}
\affiliation{
School of Nano Technology and Nano Bionics, University of Science and Technology of China, Suzhou 215123, P.~R.~China}

\author{Jiandong Sun}\email[Corresponding author, ]{jdsun2008@sinano.ac.cn}
\affiliation{
Key Laboratory of Nanodevices and Applications, 
Suzhou Institute of Nano-tech and Nano-bionics (SINANO), 
Chinese Academy of Sciences, 
398 Ruoshui Road, Suzhou 215123, P.~R.~China}

\author{Zhipeng Zhang}
\affiliation{
Key Laboratory of Nanodevices and Applications, 
Suzhou Institute of Nano-tech and Nano-bionics (SINANO), 
Chinese Academy of Sciences, 
398 Ruoshui Road, Suzhou 215123, P.~R.~China}

\author{Yunfei Sun}
\affiliation{
College of Electronic and Information Engineering, 
Suzhou University of Sciences and technology, Suzhou 215009, P.~R.~China}

\author{Yao Yu}
\affiliation{
Key Laboratory of Nanodevices and Applications, 
Suzhou Institute of Nano-tech and Nano-bionics (SINANO), 
Chinese Academy of Sciences, 
398 Ruoshui Road, Suzhou 215123, P.~R.~China}
\affiliation{
Graduate University of Chinese Academy of Sciences, Beijing 100049, P.~R.~China}

\author{Xingxin Li}
\affiliation{
Key Laboratory of Nanodevices and Applications, 
Suzhou Institute of Nano-tech and Nano-bionics (SINANO), 
Chinese Academy of Sciences, 
398 Ruoshui Road, Suzhou 215123, P.~R.~China}

\author{Muchang Luo}
\affiliation{
Chongqing Institute of Optoelectronics Technology, Chongqing 400060, P.~R.~China
}

\date{\today}

\begin{abstract}
The sensitivity of direct terahertz detectors 
based on self-mixing of terahertz electromagnetic wave in field-effect transistors 
is being improved with noise-equivalent power close to that of Schottky-barrier-diode detectors. 
Here we report such detectors based on AlGaN/GaN two-dimensional electron gas at 77~K
are able to sense broadband and incoherent terahertz radiation. 
The measured photocurrent as a function of the gate voltage agrees well with the self-mixing model 
and the spectral response is mainly determined by the antenna.
A Fourier-transform spectrometer equipped with detectors 
designed for 340, 650 and 900~GHz bands 
allows for terahertz spectroscopy in a frequency range from 0.1 to 2.0~THz.
The 900~GHz detector at 77~K offers an optical sensitivity 
about $1~\mathrm{pW/\sqrt{Hz}}$ being comparable to a commercial silicon bolometer at 4.2~K. 
By further improving the sensitivity, 
room-temperature detectors would find applications 
in active/passive terahertz imaging and terahertz spectroscopy.
\end{abstract}

\maketitle

The power of nowaday terahertz imaging systems is yet greatly limited by 
the sensitivity of available terahertz detectors. 
A broadband direct detector becomes the ideal choice for passive terahertz imaging applications, 
while a heterodyne detector is more appropriate for active and narrow-band terahertz imaging 
applications~\cite{dobroiu-mst2006,mittleman-imaging,roskos-imaging2011,roskos-imaging2014}. 
On the other hand, terahertz Fourier-transform spectrometers (THz-FTS) become less popular in terahertz spectroscopy 
due to the limited detection sensitivity and 
dynamic range comparing to terahertz time-domain spectrometers (THz-TDS)~\cite{thz-TDS,FTS-TDS-compare}.  
Even a FTS equipped with a liquid-helium (4.2~K) cooled silicon bolometer becomes less competitive 
with a THz-TDS~\cite{FTS-TDS-compare}. 
Nevertheless, silicon bolometers at 4.2~K and pyroelectric detectors at room temperature 
allow for spectroscopy in a wider frequency range than a conventional THz-TDS.  
The main drawbacks of bolometric terahertz detectors at room temperature are 
the low sensitivity and the slow response speed which limit the implementation in fast imaging and spectroscopy systems. 
Hence, a compact, ultra-sensitive and broadband terahertz detector is highly desired. 
Schottky-barrier diodes (SBD) based on GaAs have been well developed and 
are widely used in terahertz imaging systems~\cite{SBD}. 
A state-of-art SBD detector integrated in a proper waveguide offers 
a noise-equivalent power (NEP) about $2-12~\mathrm{pW/\sqrt{Hz}}$ 
in a frequency range from 0.05 to 1.1~THz~\cite{VDI-SBD}. 
When integrated with a silicon lens as a quasi-optical detector, 
the NEP of a SBD detector is about $10-25~\mathrm{pW/\sqrt{Hz}}$ in a frequency range from 0.1 to 1.0~THz.
Direct terahertz detectors based on field-effect transistors (FET) or 
high-electron-mobility transistors (HEMT) are merging as an alternative ultra-sensitive terahertz detectors 
for room-temperature 
applications~\cite{dyakonov-shur-1996,teppe-apl05,roskos-FET-DET-2009,knap-FET-DET-2011,sun-antenna-apl11,sun-model-apl12,sun-symmetry-apl15,sun-sensitivity-apl12,knap-nano2013,otsuji-imaging2013,roskos-4.5THz}.
FET-based terahertz detectors now have sensitivity and response bandwidth 
both comparable to those of SBD detectors~\cite{FET-SBD-compare}.
However, the wide response bandwith of FET/HEMT detectors was so far 
measured by using tunable single-frequency/coherent continuous-wave terahertz sources~\cite{roskos-4.5THz}  
or by using a train of (sub-)picosecond/broadband terahertz pulses in a THz-TDS~\cite{teppe-apl05,FET-SBD-compare}. 
In the latter case, the Fourier components of the pulse have fixed phase relations 
and hence are coherent in the pulse duration.
Direct detection of broadband and incoherent terahertz light from 
for example blackbodies by FET/HEMT detectors hasn't been demonstrated. 

In this letter, 
we report direct detection of  broadband incoherent
terahertz emission from hot blackbodies by using 
antenna-coupled AlGaN/GaN-HEMT detectors cooled at 77~K. 
The detection mechanism is examined 
by checking both the response spectra 
and the gate-voltage dependence under illumination 
from tunable coherent terahertz sources and 
broadband incoherent terahertz light.  
Transmission imaging of various objects 
illuminated by a hot wire resistor 
and direct imaging of hot wire resistors are demonstrated. 

\begin{figure}[!htp]
\includegraphics[width=.35\textwidth]{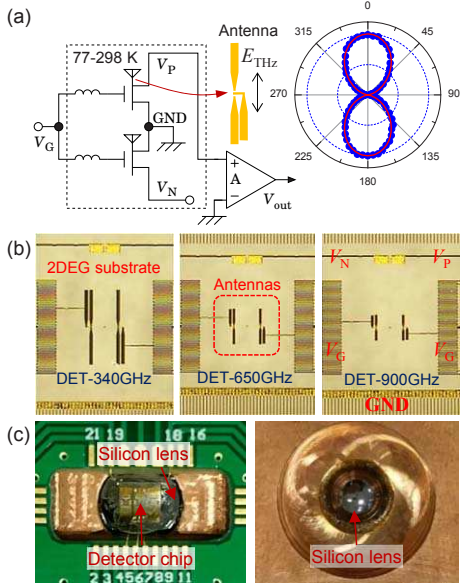}
\caption{
(a) One of the differential antenna-coupled AlGaN/GaN-HEMTs is connected 
to a low-noise voltage amplifier. 
The inset shows the polarization characteristic of the 900~GHz detector. 
(b) Partial views of detector chips designed for 340, 650 and 900~GHz bands.
(c) Backside and front-side views of the silicon hyperspherical lens 
with a detector chip assembled on the planar surface. 
}\label{fig_1}
\end{figure}

Detectors are designed based on the self-mixing mechanism~\cite{sun-antenna-apl11,sun-model-apl12,sun-symmetry-apl15} 
and fabricated using AlGaN/GaN two-dimensional electron gas (2DEG) 
on a sapphire substrate as have been reported previously~\cite{sun-model-apl12,sun-sensitivity-apl12}. 
As shown in Fig~\ref{fig_1}(a), 
the detectors are configured in a differential form, 
i.e., two HEMTs have a common source as the ground and 
two differential outputs ($V_\mathrm{P}$ and $V_\mathrm{N}$). 
The polarity of the output is determined by the asymmetric antenna. 
Since the antenna is of dipole style, 
the detector has a maximum response to terahertz wave 
with electric field polarized along the dipole direction, as shown in the inset of Fig.~\ref{fig_1}(a).  
Three detector chips named as DET-340GHz, DET-650GHz and DET-900GHz 
are fabricated with maximum response centered around 
340~GHz, 650~GHz and 900~GHz, respectively, as shown in Fig.~\ref{fig_1}(b). 
The fabrication technique and the 2DEG properties are similar to 
that reported in Ref.~\onlinecite{sun-model-apl12}. 
The 2DEG offers an electron density of $n_0=0.86\times 10^{13}~\mathrm{cm^{-2}}$ 
at 298~K and $n_0=1.10\times 10^{13}~\mathrm{cm^{-2}}$ at 77~K.  
Accordingly, the pinch-off voltage at which the electron density under a gate 
is fully depleted is $V_\mathrm{T}\approx -3.48$~V at 298~K
and $V_\mathrm{T}\approx -3.10$~V at 77~K. 
Electron mobility of $\mu=1,880~\mathrm{cm^2/Vs}$ at 298~K
is increased to $\mu=1.54\times 10^{4}~\mathrm{cm^2/Vs}$ at 77~K.  
The gate length is about $L=900~\mathrm{nm}$ and the channel width is about $4~\mathrm{\mu m}$. 
As shown in Fig.~\ref{fig_1}(c), the detector chip is assembled 
in the center of the planar surface of a high-resistivity silicon lens 
with a diameter of 6~mm and a height of 3.87~mm. 
The detectors have a noise-equivalent power about 
$30-50~\mathrm{pW/\sqrt{Hz}}$ at room temperature and 
hence doesn't show significant signal-to-noise ratio 
when applied to sense terahertz radiation from a blackbody with temperature below 1000~K. 
However, by cooling the detectors down to 77~K, 
the electron mobility is enhanced by about 8 times 
and the noise-equivalent power is reduced to below $10~\mathrm{pW/\sqrt{Hz}}$. 
Hence, experiments reported in this letter were mainly conducted  
with the detectors cooled at 77~K. 

\begin{figure}[!htp]
\includegraphics[width=.35\textwidth]{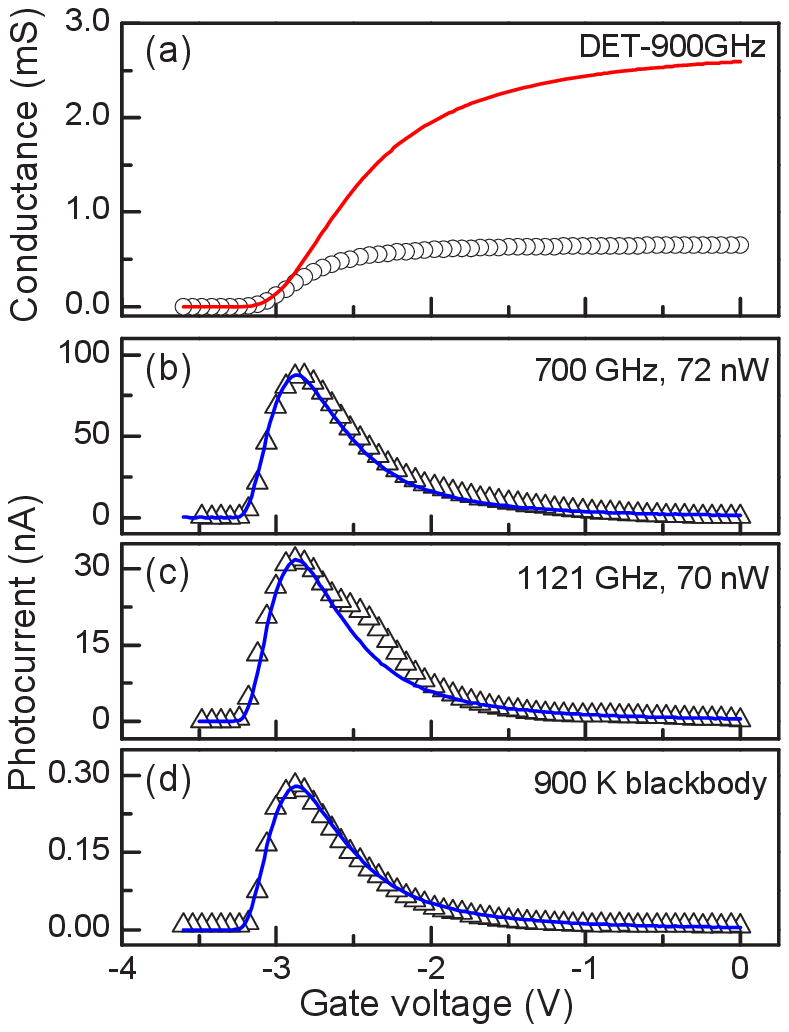}
\caption{
(a) Total conductance $G_\mathrm{m}$ of DET-900GHz at 77~K 
tuned by the gate voltage. 
The solid curve is the gate-controlled conductance $G_0$ excluding the series resistance. 
Terahertz photocurrent tuned by the gate voltage 
under continuous-wave coherent irradiation at (b) 700~GHz and (c) 1121~GHz. 
(d) Terahertz photocurrent induced by the incoherent broadband radiation 
from a blackbody with temperature about 900~K. 
The solid curves in (b), (c) and (d) are the identical simulation based on the self-mixing model.
}\label{fig_2}
\end{figure}

As shown in Fig.~\ref{fig_2}(a), 
a 900~GHz detector gives a channel conductance about $G_\mathrm{m}=0.65~\mathrm{mS}$ 
corresponding to a total resistance of $1.54~\mathrm{k\Omega}$. 
The pure gate-controlled channel conductance $G_0$ shown as the solid curve 
is extracted by excluding the large series resistance $r_\mathrm{s}\approx 1.15~\mathrm{k\Omega}$: 
$G_0=G_\mathrm{m}/(1-r_\mathrm{s}G_\mathrm{m})$.
The measured terahertz photocurrent $i_\mathrm{m}$ 
is only a fraction of the internal photocurrent $i_0$ generated 
in the gated channel: $i_\mathrm{m}=i_0(1-r_\mathrm{s}G_\mathrm{m})$. 

\begin{figure*}[ht]
\includegraphics[width=.7\textwidth]{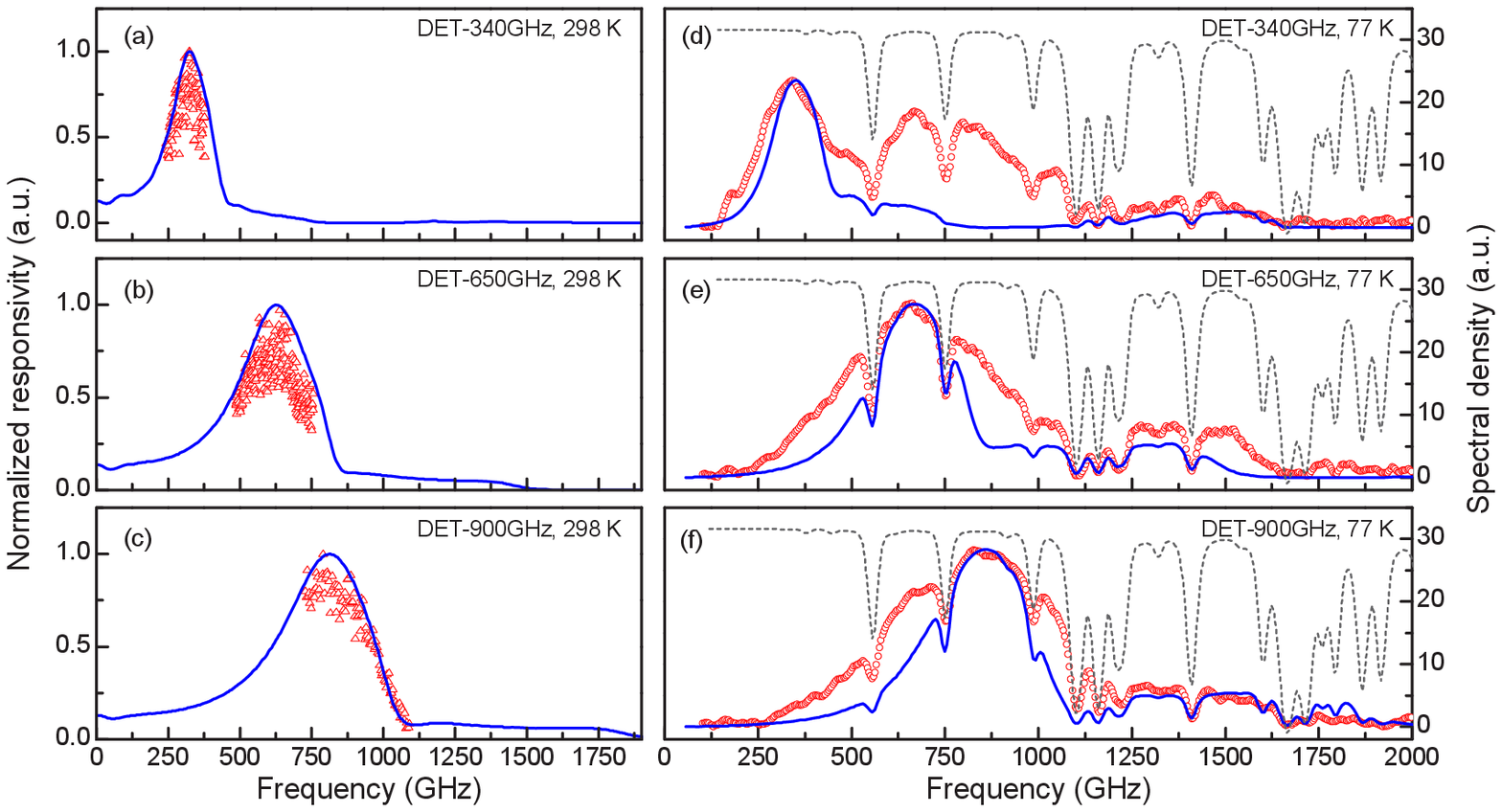}
\caption{
(a)-(c) Normalized detector responsivity for detectors at 298~K 
by using continuously tunable coherent terahertz sources. 
The solid curves are the simulated self-mixing factors. 
(d)-(f) Measured radiation spectra 
of a wire resistor at 900~K as the incoherent broadband source by detectors at 77~K. 
The solid curves are the simulated spectra and 
the dashed curves are the terahertz transmission coefficient in air.
}\label{fig_3}
\end{figure*}

According to the self-mixing model with no source-drain bias applied~\cite{sun-model-apl12}, 
the internal photocurrent can be expressed as 
\begin{equation}
i_0 = Z_\mathrm{V}\varXi(V_\mathrm{G})M(\omega)/r_0, \label{photocurrent}
\end{equation}
where 
$\varXi(V_\mathrm{G})=\mathrm{d}G_0/\mathrm{V_\mathrm{G}}$ is the field-effect factor, 
$M(\omega)$ is the self-mixing term mainly determined by the antenna design, 
$Z_\mathrm{V}=377~\mathrm{\Omega}$ 
is the characteristic impedance of electromagnetic wave in vacuum 
and $r_0=1/G_0$ is the channel resistance. 
The self-mixing term is the time average of the product of 
the terahertz field component along the channel and 
the field component perpenicular to channel 
\begin{equation}
M(\omega) = \langle \int_0^L E_x(x,t)E_z(x,t)~\mathrm{d}x\rangle_t. \label{mixing}  
\end{equation}

Under irradiation of a single-frequency ($\omega$) coherent terahertz light, 
the self-mixing term can be written as 
$M(\omega)=S_\mathrm{A}^{-1}\varLambda(\omega)\times I(\omega)S_\mathrm{A}$ 
where $I(\omega)=E_0^2(\omega)/2Z_\mathrm{V}$ 
is the Poynting flux density of the terahertz source,  
$S_\mathrm{A}$ is effective detector area, 
$I(\omega)S_\mathrm{A}$ represents the total energy sensed by the detector, 
$S_\mathrm{A}^{-1}\varLambda(\omega)$ is the self-mixing factor 
normalized by the incident flux density and detector area. 
The self-mixing factor normalized by the flux density can be expressed as
\begin{equation}
\varLambda(\omega)=\int_0^L\bar z\dot\xi_x\dot\xi_z\cos{\phi}~\mathrm{d}x,   
\end{equation}
where $\dot\xi_x=E_x/E_0$ and $\dot\xi_z=E_z/E_0$ 
are the unitless field enhancement factors, 
$\bar z$ is the equivalent distance between the channel and the gate, 
$\phi$ is the phase difference between 
the in-plane field and the perpendicular field induced by the antenna. 
Note that phase $\phi$ is independent of time and 
varies from 0 at the $x=0$ to $\pi$ at $x=L$ determined by our antenna design~\cite{sun-model-apl12}.

Under irradiation from an incoherent broadband terahertz source, 
the induced terahertz field components seen by the channel can be expressed as 
$E_x = E_0(\omega)\dot\xi_x\cos{[\omega t+\varphi]}$ and 
$E_z = E_0(\omega)\dot\xi_z\cos{[\omega t+\varphi+\phi]}$, 
where  
$\varphi$ is the random phase of each frequency component of the incoherent radiation. 
By substituting the incoherent electric field components into Eq.~\ref{mixing}, 
the time-averaged self-mixing factor can be expressed as 
\begin{equation}
M = \int_0^{+\infty} I(\omega)\varLambda(\omega)~\mathrm{d} \omega. \label{int-mixing}
\end{equation}
The form of photocurrent can be expressed by Eq.~\ref{photocurrent}. 

Based on the above analysis, 
the characteristic of terahertz photocurrent proportional to 
the field-effect factor and the self-mixing factor 
can be used as a verification of the self-mixing mechanism.  
As shown in Fig.~\ref{fig_2}(b) and (c), 
photocurrent tuned by the gate voltage is measured under continuous-wave coherent 
terahertz irradiation with frequency set at 700~GHz and 1121~GHz, respectively. 
The maximum photocurrent occurs at an optimal 
gate voltage of $-2.86~\mathrm{V}$ 
at which the field-effect factor is maximized ($\varXi_\mathrm{max}=2.44~\mathrm{mS/V}$). 
The solid curves in Fig.~\ref{fig_2}(b) and (c) 
are fits based on the self-mixing model and 
agree well with the experiment data. 
The extra shoulder at $-2.4~\mathrm{V}$ in Fig.~\ref{fig_2}(c) 
disappears at 298~K and 
is attributed to the resonant excitation of plasma wave at 1121~GHz 
as has been oberved previously~\cite{sun-symmetry-apl15}. 
Under an incoherent broadband terahertz irradiation from 
a blackbody with temperature about 900~K, 
the photocurrent tuned by the gate voltage is plotted in Fig.~\ref{fig_2}(d).  
The photocurrent proportional to the field-effect factor $\varXi$ 
can also be well described by the same form of calculated photocurrent 
as shown in Fig.~\ref{fig_2}(b) and (c).

The detector spectral response to 
coherent continuous-wave terahertz light with different frequency 
is examined at 298~K by using a continuously tunable terahertz source. 
In the measurement, 
each detector is set with the optimal gate voltage.  
As shown in Fig.~\ref{fig_3}(a)-(c), 
the normalized spectral responsivity agrees well 
with the corresponding simulated self-mixing factor which is determined by 
the antenna geometry. 
By cooling the detectors at 77~K to detect the interferogram from 
a terahertz Fourier-transform spectrometer, 
the detector spectral response to incoherent and broadband terahertz irradiation 
is further examined.
The incoherent broadband terahertz light comes from 
a 1-$\Omega$ wire resistor with a temperature about 900~K heated by a current. 
The measured blackbody spectra are shown in Fig.~\ref{fig_3}(d)-(f).  
Also, the product of the self-mixing factor $\varLambda(\omega)$, 
the blackbody spectrum $I(\omega)$ and 
the terahertz transmission coefficient $T_\mathrm{air}(\omega)$ in air ~\cite{am}
is simulated for each detector and is plotted as the solid curve accordingly.  
The center frequencies and overall spectral shapes 
of the measured response spectra agree well with the simulations. 
The measured spectra also reveal 
clearly the fine absorption lines by water vapor in air. 
As for a reference, 
the terahertz transmission coefficient in air is plotted as the dashed curves. 
The detector has a maximum response to 
incoherent radiation when the polarization is along the dipole, 
as shown in the set of Fig.~\ref{fig_1}(a). 
Discrepancies between the measured and the simulated spectra are also visible. 
The measured spectra at 298~K can be fitted 
by the simulations much better than those obtained at 77~K.  
The response spectra at 77~K are clearly broadened. 
There are a few possible causes such as 
a non-trivial change in the antenna impedance due to the increase in electron mobility 
and the rachet photocurrent from Seebeck thermoelectric effect~\cite{popov-prb15,olbrich-prb16}. 
The observed photocurrent under incoherent broadband irradiation is indeed a kind of rachet effect.  
The antenna simulations we performed didn't take into account the 2DEG under the antenna. 
In the future, we would conduct more realistic simulations 
to examine the effect of the underlying conducting 2DEG sheet on the self-mixing factor. 

\begin{figure}[!b]
\includegraphics[width=.35\textwidth]{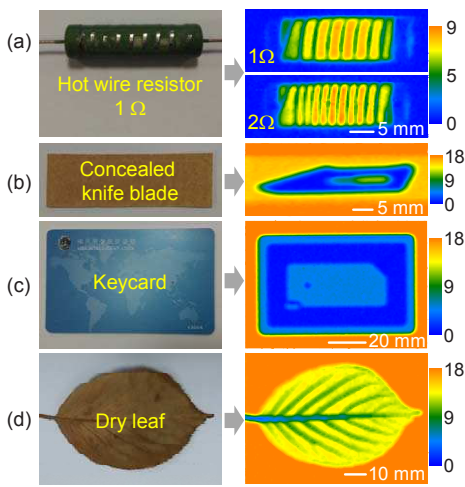}
\caption{
(a) Hot wire resistors imaged by DET-900GHz in a raster scan. 
(b) A knife blade concealed in an envelope, 
(c) a keycard and (d) a dry leaf 
are imaged by DET-900GHz in raster scans under illumination from a hot resistor.  
}\label{fig_4}
\end{figure}

Incoherent broadband terahertz emission 
from wire resistors with temperature about 900~K heated by electrical current 
is sensed by DET-900GHz and the radiation pattern  
is mapped by raster scanning the resistor 
in the focal plane of an off-axis parabolic mirror
and by fixing the detector at the focal point 
of the other off-axis parabolic mirror. 
As shown in Fig.~\ref{fig_4}(a), 
images of a 1-$\mathrm{\Omega}$ and a 2-$\mathrm{\Omega}$ wire resistors 
present clear temperature pattern according to 
the wire around the resistor body. 
Terahertz transmission imaging is demonstrated 
by using such hot wire resistor as an illumination source. 
As shown in Fig.~\ref{fig_4}(b), (c) and (d), 
a knife blade concealed in an envelope, 
a keycard and a dry leaf are raster scanned 
and the terahertz images reveal clear details of the corresponding objects. 

\begin{figure}[!htp]
\includegraphics[width=.4\textwidth]{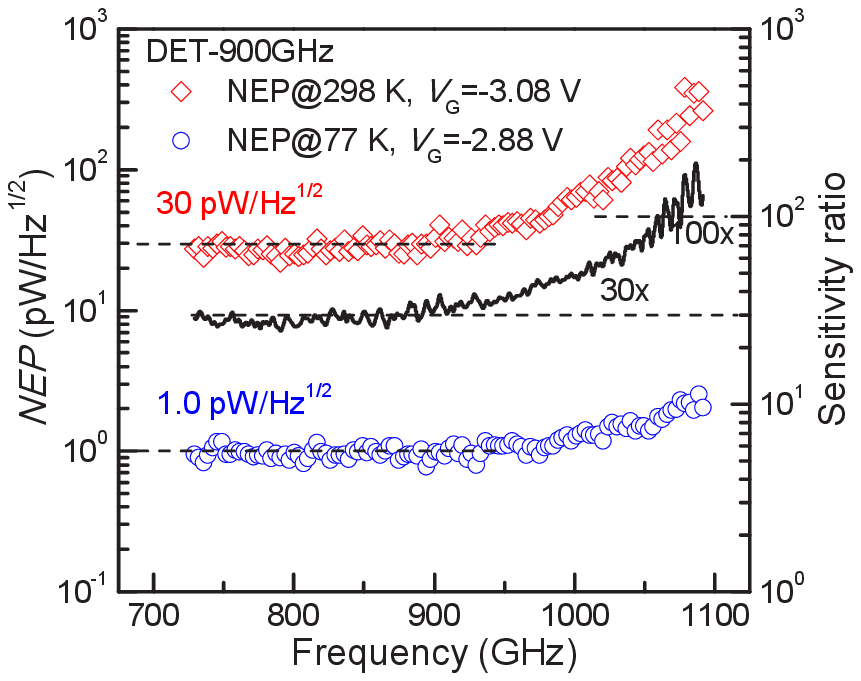}
\caption{
Different NEPs of DET-900GHz at 298~K and 77~K. 
The ratio of NEP at 298~K to that at 77~K is plotted to the right axis. 
}\label{fig_5}
\end{figure}

The optical NEP of DET-900GHz is calibrated and compared at 298~K and 77~K. 
For calibration, 
a continuous-wave coherent terahertz radiation 
is collimated and focused into a Golay-cell detector 
and the total power 
is measured at different frequencies. 
The same focused terahertz light is 
directed to the detector and the photocurrent current 
is maximized by setting the optimal gate voltage. 
As shown in Fig.~\ref{fig_5}, 
an NEP of $30~\mathrm{pW/\sqrt{Hz}}$ is achieved 
in a frequency range from 700~GHz to 925~GHz at 298~K 
corresponding to a bandwidth of $B \approx 200$~GHz. 
At 77~K, the NEP is reduced by a factor of 30 to 
about $1~\mathrm{pW/\sqrt{Hz}}$ in the same frequency range. 
The detector's optical NEP becomes comparable with that of SBD detectors at room temperature 
and that of a commercialized silicon bolometer at 4.2~K. 
The sensitivity enhancement factor by lowering the temperature 
can be as large as 100 with frequency above 1060~GHz. 
To allow for passive imaging of human bodies with 
a temperature sensitivity ($\Delta T$) better than 1~K, 
the detector's NEP needs to be in an order of $NEP\sim k_\mathrm{B}\Delta T B \sim 10^{-2}~\mathrm{pW/\sqrt{Hz}}$, 
i.e., at least three orders of magnitude to be improved  
for current room-temperature detectors. 
Recently, we have realized AlGaN/GaN-HEMT detectors with 
NEP below $3~\mathrm{pW/\sqrt{Hz}}$ at room temperature 
and there are yet two orders of magnitude to be pursued. 

In summary, 
we demonstrate detection of incoherent broadband 
terahertz radiation from hot blackbodies by using 
antenna-coupled AlGaN/GaN-HEMT detectors cooled at 77~K. 
Equipped with a Fourier-transform spectrometer, 
such detectors allow for terahertz spectroscopy in a 
wide frequency range from 0.1 to 2.0~THz. 
The spectral response and the 
responsivity tuned by the gate voltage 
are examined by using coherent and incoherent terahertz sources.  
The detector characteristics can be well 
described by the self-mixing model although 
differences are observed in response spectra 
when operated at 298~K and at 77~K. 
By further improving the sensitivity, 
AlGaN/GaN-HEMT direct terahertz detectors 
would find applications in active/passive terahertz imaging 
and terahertz spectroscopy.

\begin{acknowledgments}
The authors acknowledge supports from 
the China National Natural Science Foundation (61271157, 61401456, 61401297, 61611530708), 
the National Key Research and Development Program of China (2016YFF0100501), 
the Jiangsu Science Foundation Fund (BK20140283) and the Youth Innovation Promotion Association CAS (2017372).
\end{acknowledgments}


\begin{thebibliography}{99}

\bibitem{dobroiu-mst2006}
A. Dobroiu, C. Otani, and K. Kawase,
Meas. Sci. Technol. {\bf 17}, R161(2006). 

\bibitem{mittleman-imaging}
W. L. Chan, J. Deibel, and D. M. Mittleman, 
Rep. Prog. Phys. {\bf 70}, 1325(2007).

\bibitem{roskos-imaging2011} 
F. Friederich, W. von Spiegel, M. Bauer, F. Meng,
M. D. Thomson, S. Boppel, A. Lisauskas, B. Hils,
V. Krozer, A. Keil, T. L\"offler, R. Henneberger, A. K. Huhn,
G. Spickermann, P. H. Bol\'{\i}var, and H. G. Roskos, 
IEEE Trans. on Terahz. Sci. and Techn. {\bf 1}, 183(2011). 

\bibitem{roskos-imaging2014}
A. Lisauskas, M. Bauer, S. Boppel, M. Mundt, 
B. Khamaisi, E. Socher, R. Venckevi\u{c}ius, L. Minkevi\u{c}ius, 
I. Ka\u{s}alynas, D. Seliuta, G. Valu\u{s}is, V. Krozer, H. G. Roskos, 
J. Infrared Milli. Terahz. Waves {\bf 35}, 63(2014).

\bibitem{thz-TDS} 
W. Withayachumnankul and M. Naftaly, 
J. Infrared Milli. Terahz. Waves {\bf 35}, 610 (2014). 


\bibitem{FTS-TDS-compare}
P. Y. Han, M. Tani, M. Usami, S. Kono, R. Kersting, and X. C. Zhang, 
J. Appl. Phys. {\bf 89}, 2357 (2001).

\bibitem{SBD} 
T. W. Crowe, R. J. Matmuch, H. P. R\"oser, W. L. Bishop, 
W. C. B. Peatman, and X. L. Liu,
Proc. of the IEEE, {\bf 80}, 1827 (1992).

\bibitem{VDI-SBD}
http://www.vadiodes.com/index.php/en/products-6/detectors.

\bibitem{dyakonov-shur-1996}
M. I. Dyakonov and M. S. Shur, 
IEEE Trans. Electron Devices {\bf 43}, 380(1996).

\bibitem{teppe-apl05}
F. Teppe, D. Veksler, V. Yu. Kachorovski, A. P. Dmitriev, X. Xie， X.-C. Zhang, W. Knap， and M. S. Shur， 
Appl. Phys. Lett. {\bf 87}, 022102 (2005).

\bibitem{roskos-FET-DET-2009}
E. \"Ojefors, A. Lisauskas, D. Glaab, H. G. Roskos, U. R. Pfeiffer, 
J. Infrared Milli. Terahz. Waves {\bf 30}, 1269 (2009).

\bibitem{knap-FET-DET-2011}
F. Schuster, D. Coquillat, H. Videlier, M. Sakowicz, 
F. Teppe, L. Dussopt, B. Giffard, T. Skotnicki, and W. Knap, 
Optics Express {\bf 19}, 7827(2011). 

\bibitem{sun-antenna-apl11}
Y. F. Sun, J. D. Sun, Y. Zhou, R. B. Tan, C. H. Zeng, W. Xue, H. Qin, B. S. Zhang, and D. M. Wu, 
Appl. Phys. Lett. {\bf 98}, 252103 (2011). 

\bibitem{sun-model-apl12}
J. D. Sun, H. Qin, R. A. Lewis, Y. F. Sun, X. Y. Zhang, Y. Cai, D. M. Wu, and B. S. Zhang, 
Appl. Phys. Lett. {\bf 100}, 173513 (2012).

\bibitem{sun-symmetry-apl15}
J. D. Sun, H. Qin, R. A. Lewis, X. X. Yang, Y. F. Sun, Z. P. Zhang, X. X. Li, X. Y. Zhang, Y. Cai, D. M. Wu, and B. S. Zhang, 
Appl. Phys. Lett. {\bf 106}, 031119 (2015).

\bibitem{sun-sensitivity-apl12}
J. D. Sun, Y. F. Sun, D. M. Wu, Y. Cai, H. Qin, and B. S. Zhang, 
Appl. Phys. Lett. {\bf 100}, 013506 (2012). 

\bibitem{knap-nano2013}
W. Knap, S. Rumyantsev, M. S. Vitiello, D. Coquillat, S. Blin,
N. Dyakonova, M. Shur, F. Teppe, A. Tredicucci, and T. Nagatsuma,
Nanotech. {\bf 24}, 214002 (2013).

\bibitem{otsuji-imaging2013}
T. Watanabe, S. A. Boubanga-Tombet, Y. Tanimoto, 
D. Fateev, V. Popov, D. Coquillat, W. Knap, Y. M. Meziani, 
Y. Wang, H. Minamide, H. Ito, and T. Otsuji, 
IEEE Sensors Journal {\bf 13}, 89 (2013).

\bibitem{roskos-4.5THz}
M. Bauer, R. Venckevi\u{c}ius, I. Ka\u{s}alynas, S. Boppel, M. Mundt, 
L. Minkevi\u{c}ius, A. Lisauskas, G. Valu\u{s}is, V. Krozer, and H. G. Roskos, 
Optics Express {\bf 22}, 19250(2014). 

\bibitem{FET-SBD-compare}
S. Preu, S. Regensburger, S. Kim, M. Mittendorff, S. Winnerl, 
S. Malzer, H. Lu, P. G. Burke, A. C. Gossard, 
H.B. Weber, and M.S. Sherwin, 
Proc. of SPIE {\bf 8900}, 89000R (2013).

\bibitem{am}
The AM atmospheric model. Smithsonian Astrophysical Observatory, 
http://www.fa.harvard.edu/~spaine/am. 

\bibitem{popov-prb15}
V. V. Popov, D. V. Fateev, E. L. Ivchenko, and S. D. Ganichev, 
Phys. Rev. B {\bf 91}, 235436 (2015).

\bibitem{olbrich-prb16}
P. Olbrich, J. Kamann, M. K\"onig, J. Munzert, L. Tutsch, J. Eroms, 
D. Weiss, Ming-Hao Liu, L. E. Golub, E. L. Ivchenko, V. V. Popov, 
D. V. Fateev, K. V. Mashinsky, F. Fromm, Th. Seyller, and S. D. Ganichev,
Phys. Rev. B {\bf 93}, 075422 (2016).

\end{thebibliography}
\end{document}